\documentclass[structabstract]{aa}  
\usepackage{graphicx}
\usepackage{natbib}
\newcommand{\HI}{H\,{\sc i}}
\newcommand{\HII}{H\,{\sc ii}}
\newcommand{\Ha}{H$\alpha$}
\newcommand{\skms}{\ensuremath{\,\mbox{km}\,\mbox{s}^{-1}}}
\newcommand{\kms}{\ensuremath{\mbox{km}\,\mbox{s}^{-1}}}

\newcommand{\vsys}{\ensuremath{v_{\rm sys}}}

\newcommand{\vrot}{\ensuremath{v_{\rm rot}}}

\newcommand{\Msun}{~M$_{\odot}$}

\usepackage{txfonts}
\begin{document}
   \title{A kinematic study of the irregular dwarf galaxy NGC\,4861 using \HI\
   and \Ha\ observations}


   \author{J. van Eymeren
          \inst{1,2,3}
	  \and
          M. Marcelin\inst{4}
	  \and
	  B.~S. Koribalski\inst{3}
	  \and
	  R.-J. Dettmar\inst{2}
	  \and
	  D.~J. Bomans\inst{2}
	  \and
	  J.-L. Gach\inst{4}
	  \and
	  P. Balard\inst{4}
          }

   \offprints{J. van Eymeren}

   \institute{Jodrell Bank Centre for Astrophysics, School of Physics \&
              Astronomy, The University of Manchester, Alan Turing Building,
              Oxford Road, Manchester, M13 9PL, UK\\
              \email{Janine.VanEymeren@manchester.ac.uk}
              \and
	      Astronomisches Institut der Ruhr-Universit\"at Bochum,
              Universit\"atsstra{\ss}e 150, 44780 Bochum, Germany
	      \and
	      Australia Telescope National Facility, CSIRO,
              P.O. Box 76, Epping, NSW 1710, Australia
              \and 
             Laboratoire d'Astrophysique de Marseille, OAMP, Universit\'e
              Aix-Marseille \& CNRS, 38 rue Fr\'ed\'eric Joliot-Curie, 13013
              Marseille, France
             }

   \date{Accepted 20 July 2009}

 
  \abstract
   {Outflows powered by the injection of kinetic energy from massive stars can
  strongly affect the chemical evolution of galaxies, in particular of dwarf
  galaxies, as their lower gravitational potentials enhance the chance of a
  galactic wind.}
   {We therefore performed a detailed kinematic analysis of the neutral and
  ionised gas components in the nearby star-forming irregular dwarf galaxy
  NGC\,4861. Similar to a recently published study of NGC\,2366, we want to
  make predictions about the fate of the gas and to discuss some general
  issues about this galaxy.}
   {Fabry-Perot interferometric data centred on the \Ha\ line were obtained
  with the 1.93m telescope at the Observatoire de Haute-Provence. They were
  complemented by \HI\ synthesis data from the VLA. We performed a Gaussian
  decomposition of both the \Ha\ and the \HI\ emission lines in order to
  search for multiple components indicating outflowing gas. The expansion
  velocities of the detected outflows were compared to the escape velocity of
  NGC\,4861, which was modelled with a pseudo-isothermal halo.}
   {Both in \Ha\ and \HI\ the galaxy shows several outflows, three directly
  connected to the disc and probably forming the edges of a supergiant shell,
  and one at kpc-distance from the disc. We measured velocity offsets of 20 to
  30\skms, which are low in comparison to the escape velocity of the galaxy
  and therefore minimise the chance of a galactic wind.}
   {}

   \maketitle
%

\section{Introduction}
\label{N4861intro}
The low metal content of dwarf galaxies and the metal enrichment of the
intergalactic medium (IGM) both suggest that mass loss triggered by star
formation activity has to play a major role in the evolution of galaxies,
especially of dwarf galaxies. These galaxies provide a perfect environment to
study feedback processes as they show strong starburst activity, but are simple
systems, which makes the interaction between stars and the interstellar medium
(ISM) very efficient. Their gravitational potential is low, which supports the
long-term survival of shells, filaments, and holes. This results in numerous
ionised gas structures, sometimes up to kpc distances away from any place of
current star formation \citep{Hunter1993}, visible on deep \Ha\
images. High-resolution long-slit echelle spectra revealed that most of them
expand from the disc into the halo of their host galaxy
\citep[e.g.,][]{Martin1998,vanEymeren2007}, probably driven by stellar winds
of, e.g., Wolf-Rayet stars and by supernovae explosions.

The final fate of the expanding gas is still a matter of discussion. The
expansion into a very dense medium or a decrease of energy input could lead to
a decrease of the expansion velocity so that the gas will eventually fall back
onto the galactic disc (outflow). On the other hand, it is also possible that
the energy input is high enough to accelerate the gas beyond the escape
velocity of the host galaxy, which means that the gas might leave the
gravitational potential by becoming a freely flowing wind (galactic wind).

So far, no clear evidence for a galactic wind in local dwarf galaxies has
been found \citep{Bomans2005}. This is supported by 2D calculations of
multi-supernova remnants evolving in dwarf galaxies \citep{Silich1998}, which
show that galaxies with an ISM mass of the order of 10$^9$\Msun\ keep their
processed material, even their metals. However, hydrodynamic simulations by
\citet{MacLow1999} reveal that at least the metals always have a high chance
of being blown away, independent of the galaxy's mass. For low mass dwarf
galaxies ($<10^9$\Msun), the chance rises that also some parts of the gas
escape from the gravitational potential.

This is the second paper where we present the results of a detailed kinematic
study of the neutral and ionised gas components in a sample of nearby dwarf
galaxies. It is focused on the irregular dwarf galaxy NGC\,4861. For our
optical observations, we used a scanning Fabry-Perot interferometer centred on
the \Ha\ line that provides us with a complete spatial coverage of the galaxy
and relevant spectral information. For the \HI\ analysis, we combined the VLA
D array data by \citet{Wilcots1996} with VLA C array data. We added another
VLA C array data set published by \citet{Thuan2004} to improve the
\emph{uv}-coverage and the sensitivity.

NGC\,4861 is classified as an SB(s)m galaxy, although it shows no evidence for
spiral structure \citep{Wilcots1996}. Its appearance in \Ha\ is dominated by
the Giant Extragalactic \HII\ region (GEHR) I\,Zw\,49 in the south-west at
12$\rm ^h$ 59$\rm ^m$ 00.4$\rm ^s$, +34\degr\ 50\arcmin\ 42\arcsec, where most
of the star formation occurs. A chain of small \HII\ regions extends to the
north-east forming a tail and therefore giving the galaxy a cometary-shaped
appearance. We adopt a distance of 7.5\,Mpc \citep{deVaucouleurs1991}. 

High-resolution long-slit echelle spectra centred on
the \Ha\ line of this galaxy were recently analysed by us
\citep{vanEymeren2007}. We found a kpc-sized expanding supergiant shell (SGS4)
around the GEHR, the blue-shifted component expanding with 110\skms\ and the
red-shifted component with 60\skms. The galaxy has also been studied in \HI\ by
\citet{Wilcots1996} as well as by \citet{Thuan2004}. They report the detection
of a small \HI\ cloud, NGC\,4861\,B, at a deprojected distance of 4\,kpc east
from NGC\,4861 that appears to have no optical counterpart.

This paper is organised as follows: the observations and the data reduction
are presented in \S$\,$2. \S$\,$3 describes and compares the morphology of
both gas components; \S$\,$4 presents a kinematic analysis. In \S$\,$5,
different aspects are discussed, and \S$\,$6 summarises the main results.
%

\section{Observations and data reduction}
\subsection{The Fabry-Perot \Ha\ data}
Fabry-Perot (FP) interferometry of NGC\,4861 was performed on the 28th of
February 2006 with the 1.93m telescope at the Observatoire de Haute-Provence
(OHP), France. We used the Marseille's scanning FP and the new photon counting
camera \citep{Gach2002}. The field of view is
  5\farcm8\,$\times$\,5\farcm8 on the 512\,x\,512 pixels of the detector and
  is slightly limited by the interference filter to
  5\farcm5\,$\times$\,5\farcm5, which results in a spatial resolution of
0\farcs68 per pixel. The \Ha\ line was observed through an interference
filter centred at the galaxy's rest wavelength of 6581\,{\AA} with a Full
Width at Half Maximum (FWHM) of 10\,{\AA}. The free spectral range of the
  interferometer -- 376\skms\ -- was scanned through 24 channels with a
  sampling step of 15\skms. The final spectral resolution as measured from
the night sky lines is about 50\skms. The seeing was about 3\arcsec\ to
4\arcsec.

43 cycles were observed with an integration time of 10 sec per channel, hence
240\,sec per cycle and 172\,min in total. We used a neon lamp for the phase
and the wavelength calibration. The data reduction was done with the software
package ADHOCw\footnote{http://www.oamp.fr/adhoc/}, written by Jacques
Boulesteix: first, all cycles were checked for bad data. Afterwards, the
remaining cycles were added channel per channel. A phase map was created from
the exposure of a neon calibration lamp in order to define an ``origin''
channel, i.e., a channel where all pixels have their intensity peaking at
maximum. As this channel varies across the field of view, intensity maxima of
many pixels had to be shifted to other channels in order to get the same
wavelength origin for all pixels. As a next step, the phase map was used in
order to perform the wavelength calibration. In order to match the average
seeing at OHP, we applied a Gaussian spatial smoothing with a FWHM of 3 pixels
in x and y. Additionally, we spectrally smoothed the data with a Gaussian FWHM of 3\,channels. As a last step, we measured the night sky at different positions
across the field of view and subtracted an average value from every pixel. As
the sky intensity varies over the field of view due to the shape of the
interference filter, this approach led to residuals of the OH lines in the
spectra. However, we carefully checked their positions for the chosen free
spectral range and the neighbouring ones and conclude that the \Ha\ emission
of NGC\,4861 is not contaminated by the night sky. We also checked the
instrumental response by looking at the profile of the neon calibration
line. We found that the calibration line is symmetric. Assuming that the FP
response is similar for \Ha\ and the wavelength range where the calibration
line has been observed, we expect no artificial wings in the observed \Ha\
emission lines.

The \Ha\ intensity distribution and the velocity field were created from the
\Ha\ cube by removing all emission below a 2.5$\sigma$ threshold.
\subsection{Optical imaging}
\label{Optimage}
We use a fully reduced 900\,s \emph{R}-band image, observed with the WIYN
  3.5m telescope at Kitt Peak, which was kindly provided by
  E. Wilcots. Additionally, we work with a fully reduced 1200\,s \Ha\ image
  which is part of the ``Palomar/Las Campanas Imaging Atlas of Blue Compact
  Dwarf Galaxies'' by \citet{GildePaz2003}.

In order to search for an optical
  counterpart of the \HI\ cloud NGC\,4861\,B, we obtained a 90\,min
  \emph{V}-band image with the 2.2m telescope at Calar Alto equipped with
  CAFOS on the 3rd of March 2008. The data reduction was performed by us using
  the software package IRAF and included standard procedures of overscan- and
  bias-subtraction as well as a flatfield correction. Additionally, we removed
  cosmic rays by running the IRAF version of
  L.A. Cosmic\footnote{http://www.astro.yale.edu/dokkum/lacosmic/}
  \citep{vanDokkum2001}.
\subsection{The \HI\ data}
VLA C and D array data were kindly provided by E. Wilcots. They are flux
  and bandpass calibrated using periodic observations of 3C\,147. Further
  details concerning the data reduction can be found in \citet{Wilcots1996}. We
complemented the data by an archival VLA C array data set in order to improve
the sensitivity and the uv-coverage. Flux and bandpass calibrations using the
calibrator source 3C\,286 were done by us in AIPS. The bandpass represents the
  instrumental response. By subtracting it from the \emph{uv} data, we made
  sure that the \HI\ emission line profiles are not affected by artificial
  structures. The continuum was subtracted from the \HI\ emission by using a
  first order fit to the line-free channels.

The three data sets were then combined by taking into account the
different pointings. The \emph{invert} task in MIRIAD is very well suited for
mosaicing data so that we combined, inverted, cleaned, and restored the cube in
MIRIAD. The final data cube has a synthesised beam size of
31\arcsec$\times$\,30\arcsec\ after using a ``natural'' weighting. The spectral
resolution is 5.2\skms. We applied a 3-point Hanning smoothing, which improved
the noise level from 0.6 to 0.5\,mJy\,beam$^{-1}$.

The moment maps were created from the un-smoothed cube and again, emission
below a 2.5$\sigma$ threshold was removed. The processing and the subsequent
analysis of the \HI\ data was performed with GIPSY\footnote{The Groningen
  Image Processing System} \citep{vanderHulst1992}.
\section{General morphology}
\subsection{Stellar and ionised gas distribution}
\label{N4861morpho_opt}
\begin{figure*}
\centering
\includegraphics[width=\textwidth,viewport= 55 524 426 716,clip=]{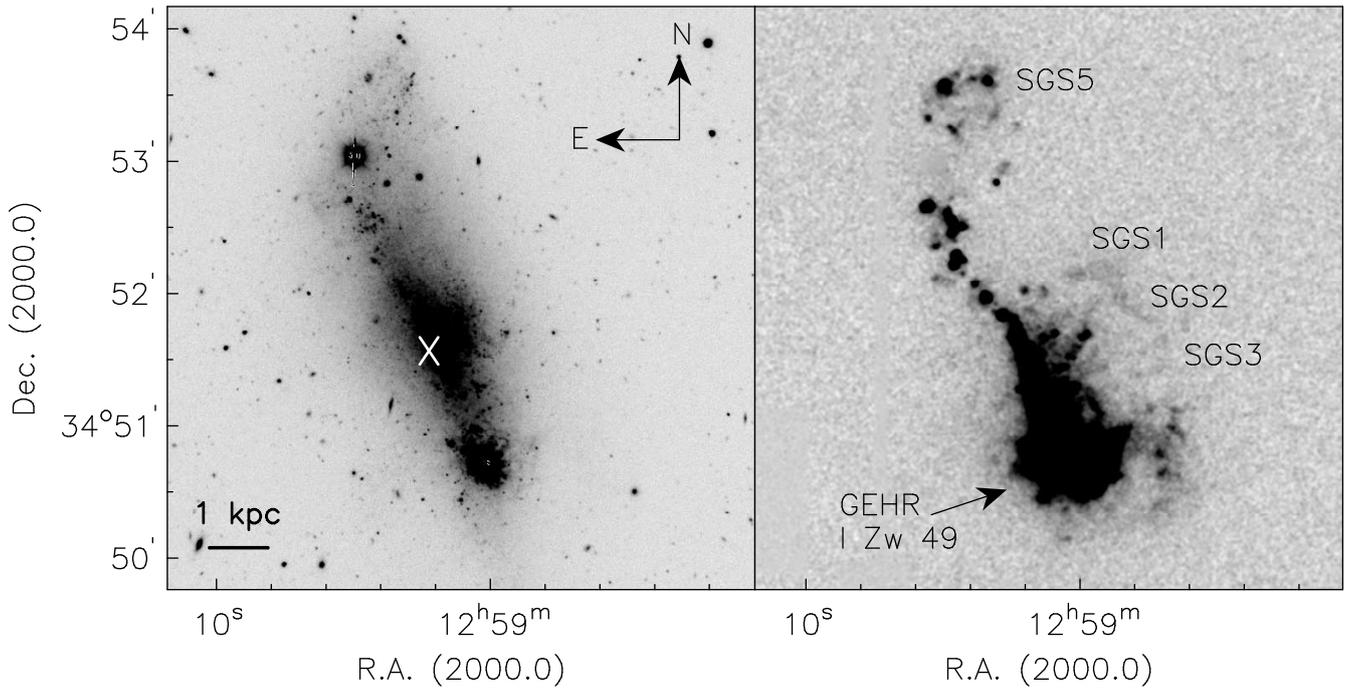}
\caption[\emph{R}-band and \Ha\ image of NGC\,4861.]{{\bf Left panel:}
  \emph{R}-band image of NGC\,4861 as obtained from the WIYN 3.5m
  telescope. The optical centre is indicated by a white cross. {\bf Right
  panel:} continuum-subtracted \Ha\ image from the Palomar/Las Campanas
  survey. The contrast is chosen in a way to emphasise the large-scale
  structures. The supergiant shells are numbered according to
  \citet{vanEymeren2007} and Sect.~\ref{N4861morpho_opt}. In order to stress
  weaker structures and to differentiate them from the noise, we used adaptive
  filters based on the H-transform algorithm \citep{Richter1991}.}
\label{N4861r+ha}
\end{figure*}

The \emph{R}-band image and the \Ha\ image from the Palomar/Las Campanas
survey are shown in Fig.~\ref{N4861r+ha}. The \Ha\ luminosity is dominated by
the GEHR in the south. A chain of smaller \HII\ regions extends to the
north-east ending in a ring-like structure at the northernmost tip, which is
located at the outer edge of the stellar distribution. No star cluster seems
to be associated with it.  With a diameter of about 1\,kpc, it belongs to the
large-scale structures in this galaxy (note that the galaxy has an optical
size of about 9\,kpc\,$\times$\,3\,kpc at the adopted distance of
7.5\,Mpc). Following the classification and numbering of
\citet{vanEymeren2007}, it is catalogued as supergiant shell SGS5.

Figure~\ref{N4861ha} shows an enlargement of the continuum-subtracted \Ha\
image that emphasises the small-scale structure close to the disc. Most of
the filaments can be found on the western side of the galaxy. Overlaid in
grey are the FP \Ha\ intensity contours at 0.5 (3$\sigma$), 5, 10, 20, 50,
100, and 300 (arbitrary units). Although the FP image is quite noisy, all
important filaments and shells can be detected.
\subsection{The distribution of the neutral gas}
\begin{figure*}
\centering
\includegraphics[width=.98\textwidth,viewport= 47 229 518 657]{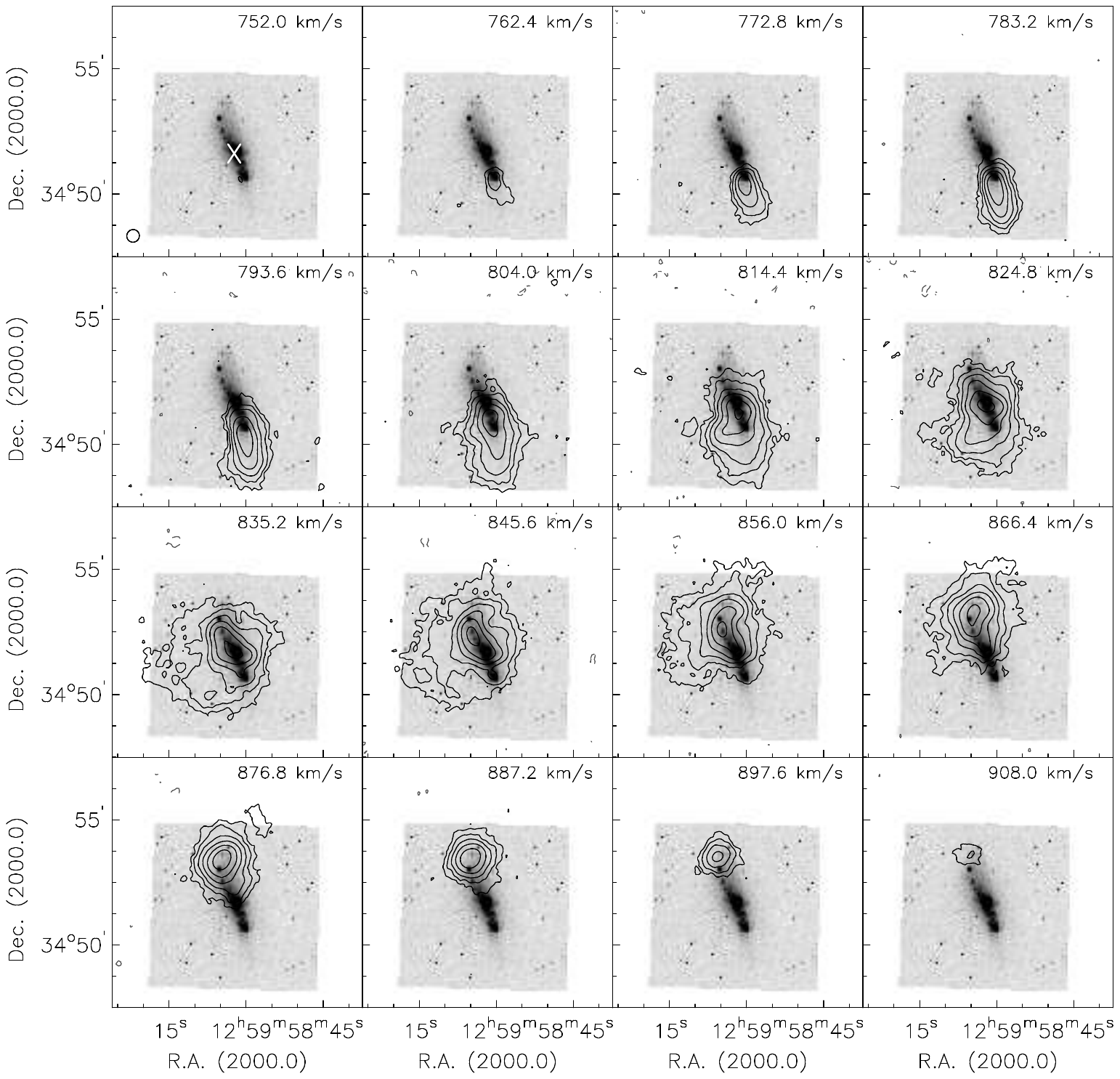}
\caption[\HI\ channel maps of NGC\,4861 (contours) superposed on the
  \emph{R}-band image.]{\HI\ channel maps of NGC\,4861 (contours) as obtained
  from the VLA using a ``natural'' weighting, superposed on the
  \emph{R}-band image. A 3-point Hanning smoothing was applied to the cube
  improving the noise level from 0.6 to 0.5\,mJy\,beam$^{-1}$. The original
  channel spacing is 5.2\skms. Contours are drawn at $-$1.5 ($-$3$\sigma$),
  1.5 (3$\sigma$), 3, 6, 12, 24 and 48\,mJy\,beam$^{-1}$. The synthesised beam
  is placed in the lower left corner of the first channel map. The optical
  centre of the galaxy is marked by a white cross in the same channel map. The
  corresponding heliocentric velocities are shown in the upper right corner of
  each channel.}
\label{N4861chan}
\end{figure*}

Figure~\ref{N4861chan} shows the \HI\ channel maps, superposed on a
greyscale presentation of the \emph{R}-band image. The white cross in the
first channel marks the optical centre of the galaxy (see
Table~\ref{n4861hiparams}). The beam is placed in the lower left corner of
the same channel. The corresponding heliocentric velocities are indicated in
the upper right corner of each channel map. At first glance, it can be seen
that the \HI\ emission is much more extended than the optical content of the
galaxy. The distribution is quite smooth with an asymmetry to the east.

\begin{figure*}
\centering
\includegraphics[width=\textwidth,viewport= 53 319 578 713,clip=]{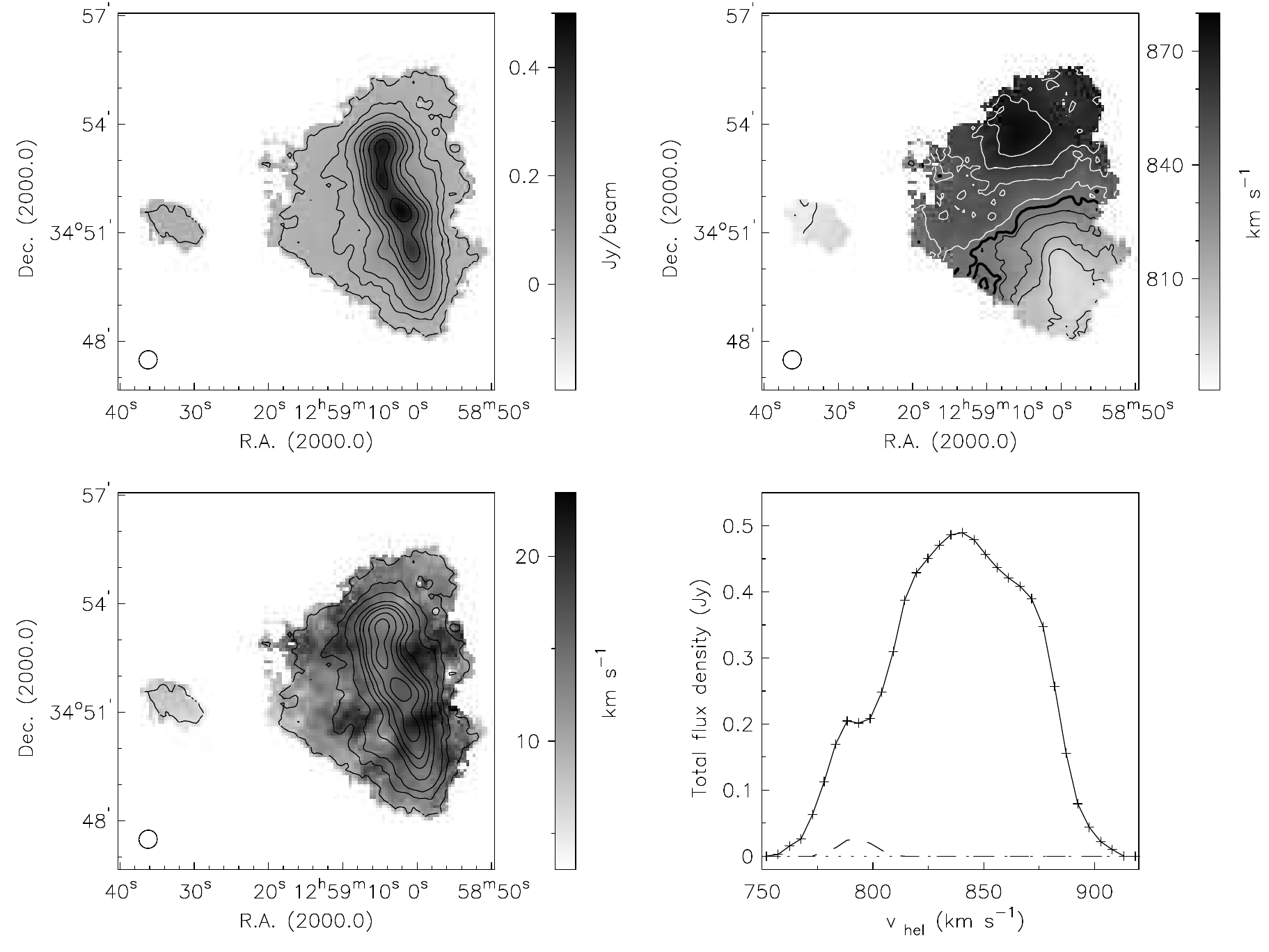}
\caption[The \HI\ moment maps of NGC\,4861.]{The \HI\ moment maps of NGC\,4861
  using ``natural'' weighting, which leads to a synthesised beam of
  31\arcsec\ $\times$ 30\arcsec. {\bf Top left:} the \HI\ intensity
  distribution. Contours are drawn at 0.01, 0.03, 0.05, 0.1, 0.15, 0.2, 0.3,
  0.4\,Jy\,beam$^{-1}$ where 10\,mJy\,beam$^{-1}$ correspond to a column
  density of $\rm 1.6\times 10^{21} atoms\,cm^{-2}$. {\bf Top right:} the \HI\
  velocity field. Contours are drawn from 790 to 870\skms\ in steps of
  10\skms. The systemic velocity of 835\skms\ is marked in bold. {\bf Bottom
  left:} the velocity dispersion, overlaid are the same \HI\ intensity
  contours as mentioned above. {\bf Bottom right:} the global intensity
  profile of the galaxy (solid line) and the \HI\ cloud NGC\,4861\,B
  (long-dashed line). The short-dashed line marks zero intensity.}
\label{N4861HI}
\end{figure*}

The integrated \HI\ intensity distribution is displayed on the upper left
panel of Fig.~\ref{N4861HI}. As already seen on the channel maps, it is
symmetrically distributed, except for the eastern part, with several maxima
along the north-south axis. The distortion in the east is most probably linked
to the \HI\ cloud NGC\,4861\,B. Its morphology will be discussed separately in
Sect.~\ref{HIcloudmorpho}.
\subsection{A comparison of the neutral and ionised gas distribution}
\label{N4861morpho}
\begin{figure}
 \centering
\includegraphics[width=.49\textwidth,viewport=55 521 250 717,clip=]{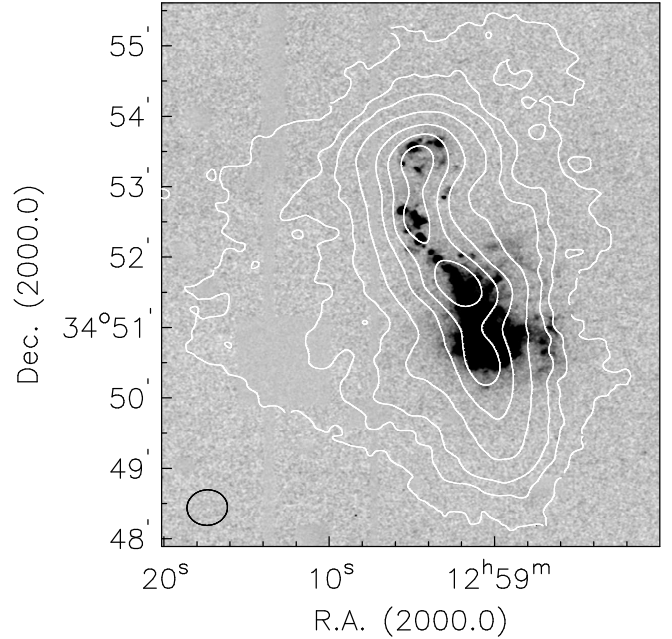}
\caption[A comparison of the \Ha\ and the \HI\ morphology.]{A comparison of
  the \Ha\ and the \HI\ morphology. The continuum-subtracted \Ha\ image is
  displayed. Overlaid in white are the \HI\ intensity contours at 0.01, 0.03, 0.05, 0.1, 0.2, 0.3, and 0.4\,Jy\,beam$^{-1}$.}
\label{N4861ha+hi}
\end{figure}

We now compare the morphology of the ionised and neutral gas by plotting the
\HI\ intensity contours over the continuum-subtracted \Ha\ image (see
Fig.~\ref{N4861ha+hi}). The overlay shows that the optical galaxy lies in
the centre of the \HI\ distribution. The \HI\ intensity maxima all coincide
with the optical extent. However, the \HI\ column density coinciding with the
GEHR is lower than the maxima along the tail. Furthermore, the \HI\ peak
intensity is by more than one beam size offset with respect to the centre of
the GEHR. The lower \HI\ column density indicates that a significant part of
the neutral gas has already been ionised, which resulted in the huge luminous
\HII\ region. Such an offset has also been found in other galaxies (IC\,10,
NGC\,2366) and has been explained by sequential star formation
\citep{Hodge1994}.

The three main \HI\ maxima coincide with several of the smaller \HII\ regions
along the tail. Close to the supergiant shells SGS1, SGS2 and SGS3 as well as
east of the GEHR, the gradient of the intensity decrease is different from the
overall gradient. In both cases, the gradient is flatter, which means that the
gas density is higher than in the surroundings. This might indicate outflowing
gas and indeed, the kinematic analysis below will show that these two regions
harbour expanding gas structures.
\subsection{Morphological features of the HI cloud NGC\,4861\,B}
\label{HIcloudmorpho}
\begin{figure*}
\centering
\includegraphics[width=.98\textwidth,viewport= 47 429 518 657]{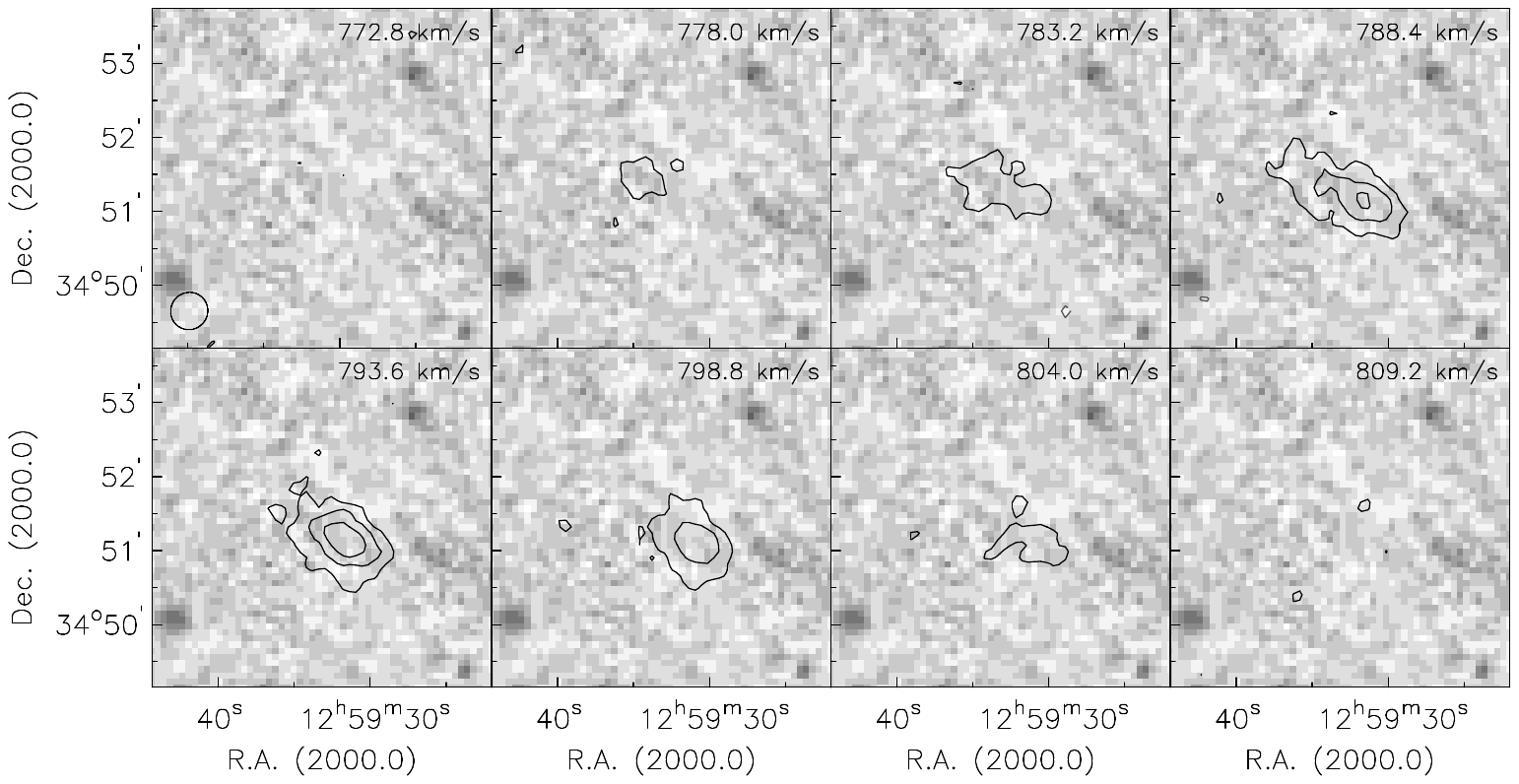}
\caption[\HI\ channel maps of NGC\,4861 B (contours) superposed on the
  \emph{V}-band image.]{\HI\ channel maps of NGC\,4861 B (contours),
  superposed on our \emph{V}-band image. No smoothing was applied. Contours
  are drawn at $-$2 ($-$3$\sigma$), 2 (3$\sigma$), 4, and
  6\,mJy\,beam$^{-1}$. Otherwise the same as Fig.~\ref{N4861chan}.}
\label{N4861chanB}
\end{figure*}

NGC\,4861 has a small companion east of the main body at a deprojected
  distance of 4\,kpc (Fig.~\ref{N4861HI}). The \HI\ mass was measured to be
  4$\times$$10^6$\,\Msun\ (see Table~\ref{n4861hiparams}). It seems to
  interact with the main \HI\ complex as the large-scale \HI\ distribution is
  extended in its direction and the velocity field is
  distorted. Figure~\ref{N4861chanB} shows the \HI\ channel maps, superposed
  on a greyscale presentation of our \emph{V} band image. NGC\,4861\,B can be
  detected over a velocity range of 26\skms, although the column density is
  quite low. The gas is smoothly distributed.

\begin{figure}
\centering
\includegraphics[width=\textwidth,viewport= 43 324 500 523,clip=]{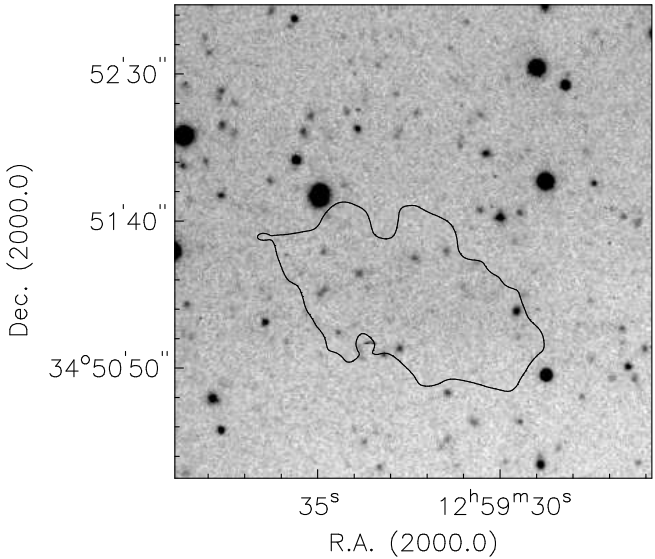}
\caption[Deep \emph{V}-band image of the area of NGC\,4861\,B.]{Deep
  \emph{V}-band image of the area of NGC\,4861\,B. Overlaid in black is
  the outer \HI\ intensity contour of the \HI\ cloud at
  0.01\,Jy\,beam$^{-1}$.}
\label{N4861B}
\end{figure}

Previous studies using DSS images could not detect an optical counterpart of
this cloud \citep[e.g.,][]{Wilcots1996}. However, we have obtained a
significantly deeper \emph{V}-band image of this field (see
Sect.~\ref{Optimage}), which is shown in Fig.~\ref{N4861B}. Overlaid are the
outer \HI\ intensity contour at 0.01\,Jy\,beam$^{-1}$. No stellar association
that could be connected to NGC\,4861\,B can be seen. Assuming that
the size of a possible stellar disc equals the \HI\ extent of
1\farcm6\,$\times$\,0\farcm9 (see Table~\ref{n4861hiparams}), we derived an
upper limit for the integrated flux in the \emph{V}-band image of
$m_V<24.3$\,mag and $M_V<-6.54$\,mag respectively, which makes the existence
of a stellar component very unlikely.
\section{Kinematic analysis}
Spectra extracted from the \Ha\ and the \HI\ data cubes show that
both emission lines are sometimes split into several components. In order to
take all gas components into account and to measure their properties,
especially their velocities, we performed a Gaussian decomposition by
interactively fitting the \Ha\ and \HI\ emission with the IRAF task
\emph{splot}. Only detections above a 2.5$\sigma$ limit were considered. All
given velocities are heliocentric velocities measured along the line of
sight.
\subsection{The \Ha\ velocity field}
\label{N4861Havelo}
\begin{figure}
\centering
\includegraphics[width=.49\textwidth,viewport= 58 314 320 713,clip=]{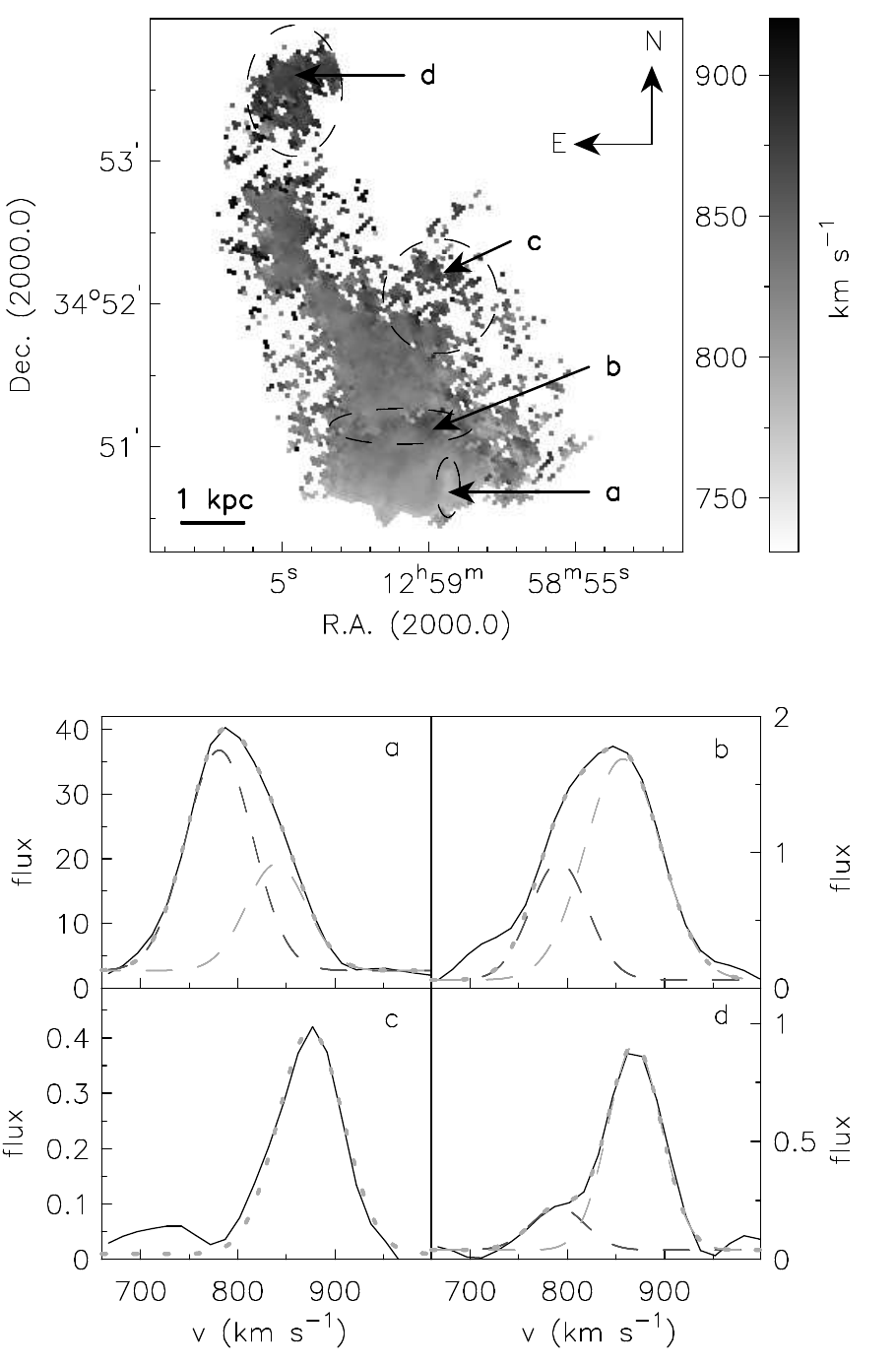}
\caption[The \Ha\ velocity field of NGC\,4861.]{The \Ha\ velocity field of
  NGC\,4861. {\bf Upper panel:} the velocity distribution of the strongest
  component. Significant deviations from the rotational gradient are
  illustrated by black ellipses. {\bf Lower panel:} four example spectra
  (black solid lines) extracted from different areas across the galaxy, which
  show the flux in arbitrary units \emph{vs.} the line of sight velocity. The
  \Ha\ line is often split into two components (blue (dark grey) long-dashed
  lines and red (light grey) long-dashed lines). The sum of both Gaussian fits
  is plotted with green (light grey) short-dashed lines. The following
  detections are displayed: {\bf (a)} the blue- and red-shifted outflows south
  of the GEHR; {\bf (b)} the blue- and red-shifted counterparts north of the
  GEHR; {\bf (c)} the red-shifted supergiant shells; {\bf (d)} the potential
  blue-shifted component at SGS5.}
\label{N4861fp}
\end{figure}

Figure~\ref{N4861fp}, upper panel shows the resulting \Ha\ velocity field with
the component of highest intensity plotted. The overall velocity gradient runs
from the south-west with velocities of about 800\skms\ to the north-east with
velocities of about 870\skms. This fits well with the systemic velocity
measured from the \HI\ data of about 835\skms\ (see below) and gives the
ionised gas a rotation velocity of 35\skms.

Several deviations from the overall rotation become immediately visible
(indicated by black ellipses in the \Ha\ velocity field, corresponding example
  spectra are shown on the lower panel of Fig.~\ref{N4861fp}):
first, we detected a strong blue- and a weaker red-shifted component in the
south of the GEHR (spectrum a, black solid line). The \Ha\ line is
very broad and asymmetric and can be well fitted with two Gaussians, one at
about 780\skms\ (blue (dark grey) long-dashed line), which is 30\skms\
blue-shifted with respect to the overall velocity gradient, and one at about
840\skms\ (red (light grey) long-dashed line), which is red-shifted by
30\skms\ with respect to the overall velocity gradient. The sum of both
Gaussian fits is plotted with a green (light grey) short-dashed line and is in
good agreement with the observed spectrum.

A strong red-shifted and a weaker blue-shifted component can be seen in the
north of the GEHR (spectrum b). Again, both gas components are expanding with
about 30\skms\ in comparison to their surroundings, which means that they
probably belong to the blue- and red-shifted gas detected in the south of the
GEHR.

Along the chain of \HII\ regions to the north-east of the galaxy,
slightly red-shifted gas seems to flow out of the disc in the direction of the
supergiant shells located in the west (see Fig.~\ref{N4861r+ha}, right
panel). These three supergiant shells are kinematically separated: SGS1 and
SGS2 are red-shifted with an expansion velocity of 30\skms\ (spectrum c),
whereas SGS3 follows the rotation velocity.

The velocities in the area of SGS5 also follow the rotation of the galaxy. We
found an indication of a blue-shifted component with a velocity offset of
about 80\skms\ with respect to the main component (spectrum d).

Note: except for the GEHR and the area of SGS5, we either detected only one
expanding component, which then dominated the spectrum, or no outflows at
all. Therefore, we only show the velocity field of the components of highest
intensity. Furthermore, the instrumental response of the FP could cause
  artificially broader \Ha\ lines \citep[e.g.,][]{Moiseev2008}. To account for
this effect, the fitted Gaussians have to be convolved with the instrumental
profile. We compared pure Gaussian fits with convolved fits and found that the
values of the emission line peaks remain unchanged, which is in agreement with
the results by \citet{Moiseev2008}. However, the \Ha\ velocity dispersion
might change significantly so that we do not mention any values in this paper.
\subsection{The \HI\ velocity field}
\label{N4861HIvelo}
The upper right panel of Fig.~\ref{N4861HI} shows the \HI\ velocity field which
looks fairly regular with a gradient rising from the south-west with
velocities of about 795\skms\ to the north-east with velocities of about
870\skms. As already mentioned by \citet{Wilcots1996}, the isovelocity
contours close at both ends of the galaxy, which is an indication for a
declining rotation curve, possibly caused by a bar. We detected two major
deviations from the rotation, the faint eastern part which does not show any
regular velocity pattern, and the small \HI\ cloud NGC\,4861\,B in the east
which is offset by about 60\skms\ with respect to the velocity of the main
body, suggesting that it is not rotating in the disc.

The \HI\ velocity dispersion map (Fig.~\ref{N4861HI}, lower left panel) shows
altogether four maxima. The two areas of high dispersion to the east are
probably caused by an interaction between the isolated \HI\ cloud and the main
body. The other two dispersion peaks might indicate an additional gas
component. The overlay of the \HI\ intensity contours reveals that the \HI\
maxima are clearly offset from the regions of high dispersion except for the
southernmost peak. All four dispersion maxima have velocities of 20 to
22\skms, the median value lies at 14\skms.
\subsection{The \HI\ rotation curve}
\label{N4861Hirot}
In order to investigate the fate of the gas, we need to know some of the
kinematic parameters of NGC\,4861 like its inclination or its rotation
velocity. Therefore, we derived a rotation curve from the \HI\ data by fitting
a tilted-ring model to the observed velocity field.

\begin{figure*}
\centering
\includegraphics[width=\textwidth,viewport= 48 324 580 713,clip=]{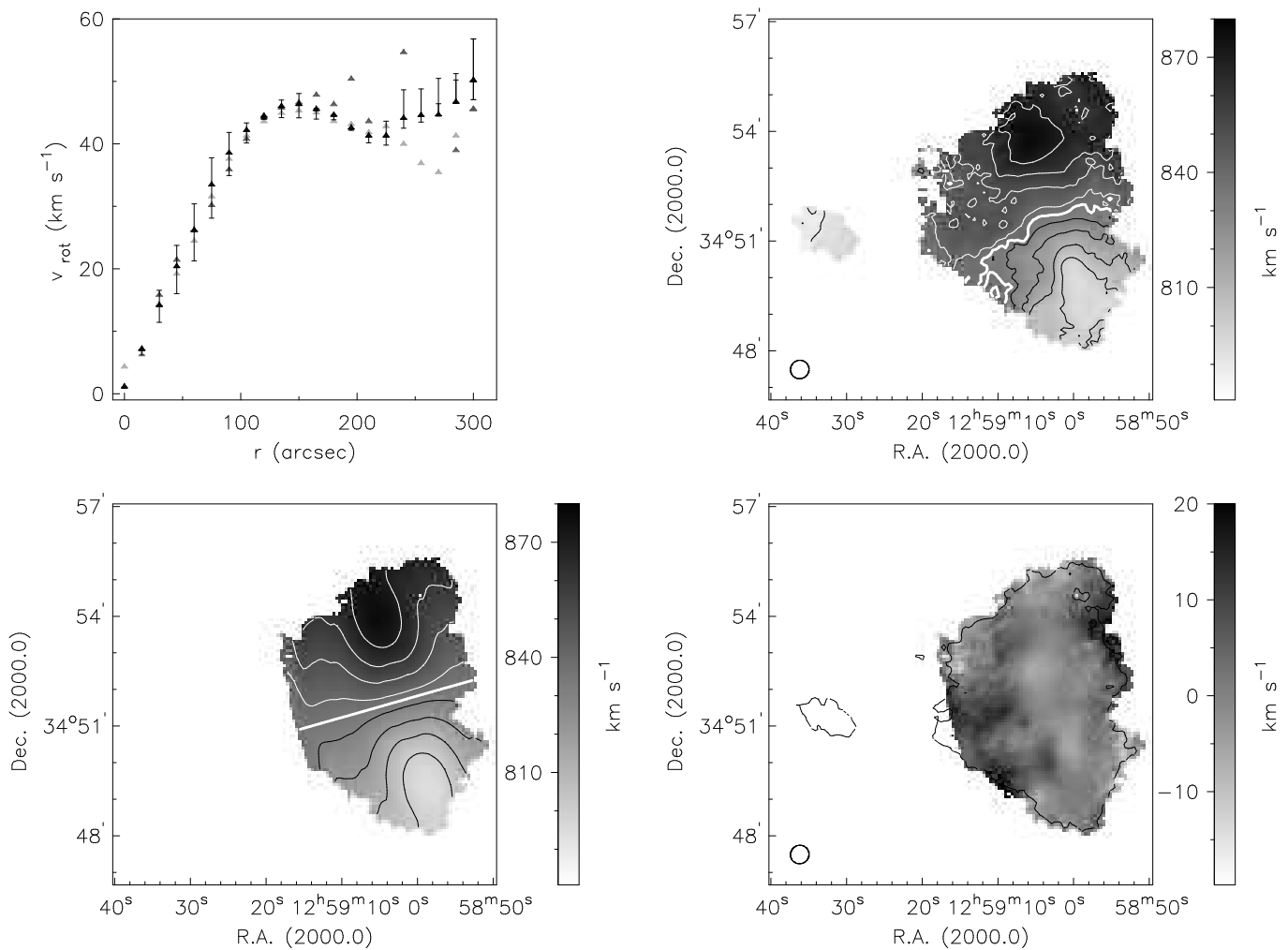}
\caption[The \HI\ rotation curve of NGC\,4861.]{The \HI\ rotation curve of
  NGC\,4861. {\bf Top left:} different approaches for deriving the rotation
  curve. The black symbols represent the best-fitting parameters derived in an
  iterative approach. The error bars indicate receding and approaching
  side. The green (light grey) curve was derived by taking the initial
  estimates and keep them fixed, the red (dark grey) curve by taking the
  best-fitting parameters and let them vary. {\bf Top right:} the \HI\
  velocity field. {\bf Bottom left:} the model velocity field, created by the
  best-fitting parameters. {\bf Bottom right:} the residual map after
  subtracting the model from the original velocity map.}
\label{N4861rotcur}
\end{figure*}

At the beginning, initial estimates for the kinematic parameters had to be
defined, which were obtained by interactively fitting ellipses to the \HI\
intensity distribution using the GIPSY task \emph{ellfit}. These were then
used as an input for the tilted-ring fitting routine \emph{rotcur}
\citep{Begeman1989}. The width
of the rings was chosen to be half the spatial resolution, i.e., 15\arcsec\ in
this case. In order to get the most precise values, three different approaches
were made by always combining receding and approaching side. First, the initial
estimates were all kept fixed. The resulting curve is indicated by the green
(light grey) symbols in Fig.~\ref{N4861rotcur}, upper left panel. In a second
approach, the parameters were iteratively defined for all rings by keeping all
parameters fixed except for the one we wanted to measure. Up to a
radius of 200\,\arcsec, which corresponds to a distance of 7\,kpc from the
dynamic centre, no significant deviation or a sudden change of any of the
parameters was noticed so that an average value (given in
Table~\ref{n4861hiparams}) was taken for each parameter (black symbols). In
this approach, a rotation curve was also measured for receding and approaching
side alone, indicated by the error bars of the black symbols. As a last
approach, the so derived parameters were all left free in order to reproduce
the result of the second approach (red (dark grey) symbols).

In the inner 200\,\arcsec, the green (light grey) symbols are in very good
agreement with the black ones. From a radius of 200\,\arcsec\ on, the
differences become larger. The reason for this is that the filling
factors of the rings drop from about 1 to about 0.5 at a radius of 200\arcsec,
which leads to a higher uncertainty in calculating the rotation
velocity. Therefore, every value above 200\,\arcsec\ has to be treated with
care. The red (dark grey) symbols only agree well within a radius of about
150\,\arcsec.

In the inner 100\,\arcsec, the velocity gradient is very steep and linear,
indicating solid body rotation, which characterises dwarf galaxies. Above
100\,\arcsec, the curve shows a plateau with a tendency to decline. The new
rise at a radius of about 240\,\arcsec\ is certainly due to the sparsely
filled tilted rings (as mentioned above). On the other hand, the disturbed gas
in the eastern parts of the galaxy probably also causes a rise in velocity.

The best-fitting parameters from the iterative approach are given in
Table~\ref{n4861hiparams}. We derived an inclination of 65\degr, a position
angle of 16\degr, and a systemic velocity of 835\skms. These values are in
good agreement with observations by \citet{Thuan2004} who measured \vsys\ to
be 833\skms\ and a position angle of 17\degr. However, their inclination of
82\degr\ is significantly higher than our value. They also derived a higher
value of \vrot\ (54\skms\ in comparison to 46\skms\ measured by us), which
cannot be explained by the difference in inclination alone. Nevertheless, our
result confirms the one of \citet{Thuan2004}, which is that the rotation
velocity measured by \citet{Wilcots1996} of 80\skms\ (\emph{i} of 67\degr) is
much too high.

In order to prove the reliability of the derived parameters, a model velocity
field with the best-fitting values was created (Fig.~\ref{N4861rotcur}, lower
left panel) and subtracted from the original velocity field (upper right
panel). The residual map can be seen on the lower right panel of
Fig.~\ref{N4861rotcur}. The overall velocity field is very well represented by
our derived parameters except for the extension in the east of the
galaxy. Here, the residuals reach absolute values of 15 to 20\skms\ in
comparison to a general value of $\pm$\,5\skms. As already mentioned before,
this extension in combination with the disturbed velocity field is likely to
be caused by an interaction with the small \HI\ cloud NGC\,4861\,B to the east
of the main body. Its kinematics are analysed in Sect.~\ref{HIcloud}.
\begin{table}
\caption{\HI\ properties as measured from the VLA data.}
\label{n4861hiparams}
$$
\begin{tabular}{lcc}
  \hline
  \hline
  \noalign{\smallskip}
  Parameters [Unit] & NGC\,4861 & NGC\,4861 B\\
  \hline
  \noalign{\smallskip}
  optical centre$^a$: & &\\
  ~~$\alpha$ (J2000.0) & 12$\rm ^h$ 59$\rm ^m$ 02.3$\rm ^s$ &\\
  ~~$\delta$ (J2000.0) & +34\degr\ 51\arcmin\ 34\arcsec&\\
  $D$ [Mpc]$^b$ & 7.5 &\\
  \hline
  \noalign{\smallskip}
  dynamic centre$^c$: & &\\
  ~~$\alpha$ (J2000.0) & 12$\rm ^h$ 59$\rm ^m$ 01.4$\rm ^s$ & 12$\rm ^h$ 59$\rm ^m$ 32.6$\rm ^s$\\
  ~~$\delta$ (J2000.0) & +34\degr\ 51\arcmin\ 43\arcsec & +34\degr\ 51\arcmin\ 15\arcsec\\
   \vsys\ [\kms]$^c$ & 835$\pm$1 & 791\\
  $i$ [\degr]$^c$ & 65$\pm$3 & 65\\
  $PA$ [\degr]$^c$ & 16$\pm$3 & 244\\
\hline
\noalign{\smallskip}
\vrot\ [\kms]$^c$ & 46 & 4.4\\
  $F_{\rm HI}$ [Jy \kms] & 36.08 & 0.30\\
  $M_{\rm HI}$ [$\rm 10^8$\,\Msun] & 4.79 & 0.04\\
  \HI\ diameter [\arcmin] & 7.0 $\rm \times$ 4.9 & 1.6 $\rm \times$ 0.9\\
  ~~~~~~~~~"~~~~~~~~~ [kpc] & 15.2 $\rm \times$ 10.7 & 3.6 $\rm \times$ 1.9\\
  \HI\ / opt. ratio & 2.3 $\rm \times$ 5& ...\\
  $<\sigma>$ [\kms] & 14 & 7\\
  $\sigma_{\rm Peak}$ [\kms] & 22.8 & 9.9\\
  $r_{\rm HI,max}$ [kpc] & 10.9 & 2.7\\
  $M_{\rm dyn}$ [$\rm 10^9$\,\Msun] & 5.4 & 0.012\\
  \noalign{\smallskip}
  \hline
\end{tabular}
$$
$^a$\,Data from NED; $^b$\,distance from \citet{deVaucouleurs1991};
  $^c$\,derived by fitting a tilted-ring model to the \HI\ data. In case
  of NGC\,4861, where all parameters were fitted iteratively, the error
  weighted mean values plus uncertainties are given. The errors of the dynamic
  centre position are far below one beam size. The parameters of NGC\,4861\,B
  were estimated from fitting ellipses to the \HI\ distribution. 
\end{table}
\subsection{Comparison of the neutral and ionised gas kinematics}
\label{Both}
\begin{figure}
\centering
\includegraphics[width=.49\textwidth,viewport= 55 522 320 713,clip=]{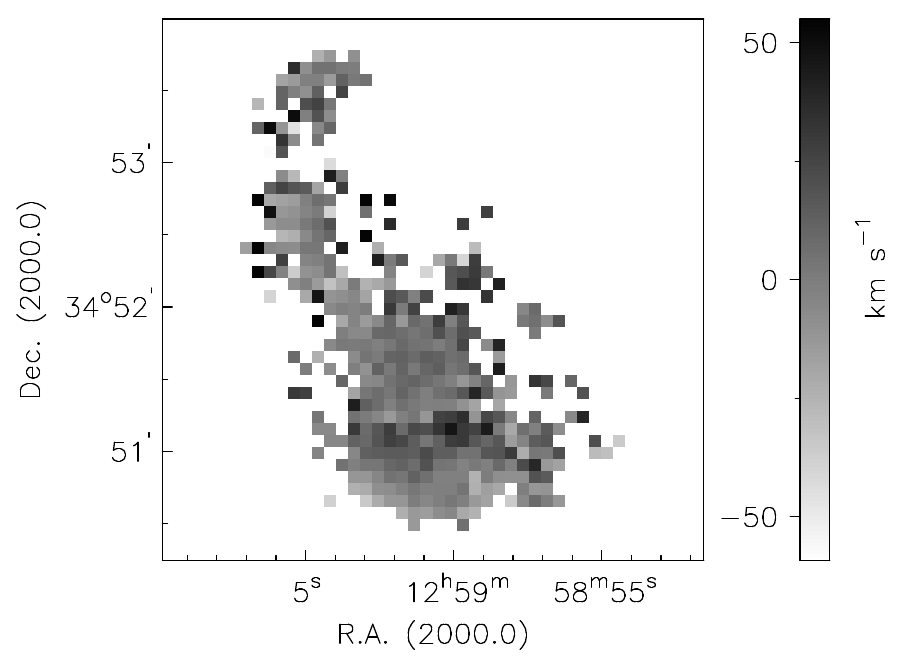}
\caption{A comparison of the neutral and ionised gas velocities. The residuals
  after subtracting the \HI\ velocity field from the \Ha\ velocity field are
  shown.}
\label{N4861all}
\end{figure}

The \Ha\ velocity field (see Fig.~\ref{N4861fp}, upper panel) shows three
major deviations from the overall rotation velocity, one in the south of the
GEHR, one north of the GEHR, and one close to SGS1. In order to compare the
velocities of the ionised gas with the velocities of the neutral gas, the \HI\
velocity field was subtracted from the \Ha\ velocity map. Therefore, the FP
data were smoothed to fit the spatial resolution of the \HI\ data of
31\arcsec$\times$\,30\arcsec. The resulting residual map is shown in
Fig.~\ref{N4861all}. At most positions, the absolute value of the residuals is
below 10\skms, which means that the velocities of the neutral and ionised gas
are generally in good agreement. The ionised gas in the southern part of the
GEHR is blue-shifted with a velocity offset of about 25\skms\ relative to
\HI. The ionised gas of SGS1 and SGS2 is red-shifted with a velocity offset of
about 30\skms. Additionally, some red-shifted gas can be seen north of the
GEHR, also with a velocity offset of about 30\skms.

\begin{figure*}
\centering
\includegraphics[width=\textwidth,viewport= 63 529 490 678,clip=]{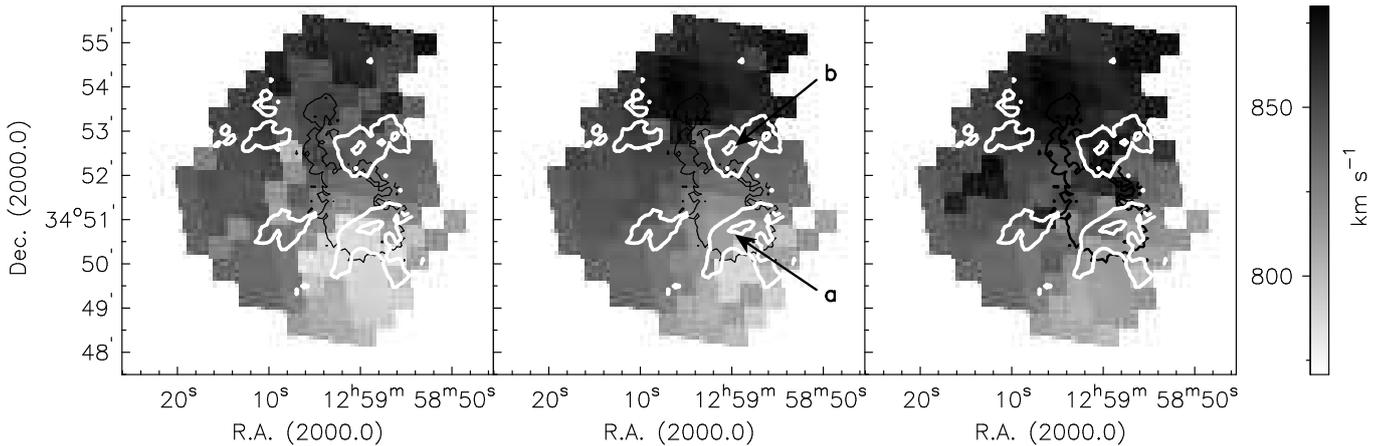}
\caption{Gaussian decomposition of the \HI\ emission line. Blue-shifted (left
  panel), main (middle panel) and red-shifted (right panel) components of the
  \HI\ velocities are shown. Note that regions where no blue- or red-shifted
  component was detected were filled with the main component. Overlaid in
  white are the \HI\ velocity dispersion contours at 18 and 22\skms\ as well
  as the outermost \Ha\ intensity contour in black. The areas from where we
  extracted example spectra are labelled.}
\label{N4861.hi.decomp}
\end{figure*}

\begin{figure}
\centering
\includegraphics[width=\textwidth,viewport=37 476 642 687,clip=]{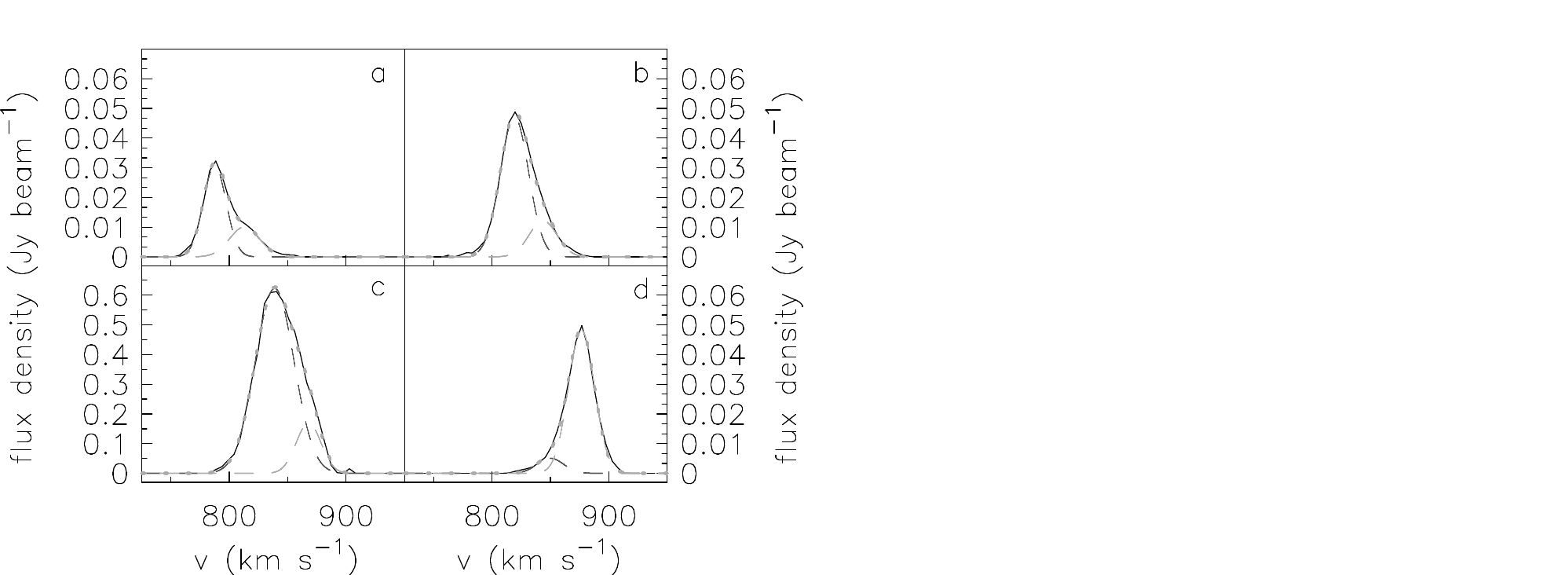}
\caption[Some example \HI\ line profiles.]{Some example \HI\ line
  profiles (black solid lines): {\bf (a)} the blue-shifted outflow south of
  the GEHR; {\bf (b)} the red-shifted counterpart north of the GEHR; {\bf (c)}
  the red-shifted outflow close to the supergiant shells; {\bf (d)} the
  potential blue-shifted component at SGS5. The Gaussian profiles fitted to
  the single components are overlaid in blue (dark grey) and red (light grey),
  the resulting sum of the profiles is plotted with a green (light grey)
  short-dashed line.}
\label{n4861lineprofileshi}
\end{figure}

The areas of high \HI\ velocity dispersion coincide fairly well with the
regions of expanding ionised gas. Therefore, we decided to have a closer look
at the kinematics of the neutral gas. We performed a Gaussian decomposition of
the \HI\ velocities by averaging the velocities over one beam size. The result
is shown in Fig.~\ref{N4861.hi.decomp}: the components of highest intensity
are presented in the middle panel. Overlaid in white are the \HI\ velocity
dispersion contours at 18 and 22\skms\ and the outer \Ha\ intensity contour in
black. For a comparison, the blue- and red-shifted components are shown on the
left and right panel respectively. Note that regions where we did not find a
blue- or red-shifted component were filled with the main component. Four
example spectra extracted from the \HI\ cube at roughly the same positions as
the \Ha\ spectra (see Fig.~\ref{N4861fp}) are displayed in
Fig.~\ref{n4861lineprofileshi} together with the Gaussian fits for the single
components (long-dashed blue (dark grey) and red (light grey) lines) and the
resulting sum (short-dashed green (light grey) lines).

Looking at the southern edge of the optical galaxy, a blue-shifted component
is detected with a heliocentric line of sight velocity of about 786\skms. The
main component at this position has a velocity of about 812\skms. The \Ha\
velocity field shows at the same position a blue-shifted outflow with
velocities of about 780 to 790\skms, which is in good agreement with the \HI\
data. Additionally, we detected a red-shifted component at about 840\skms\ in
the \Ha\ spectrum (see Fig.~\ref{N4861fp}, spectrum a). At the same position,
the Gaussian decomposition of the \HI\ profiles also reveals gas at about
837\skms, which is, however, not as extended as the blue-shifted gas. The
blue-shifted outflow has already been detected in \HI\ by \citet{Thuan2004}
who suggested that this gas had been moved away from the GEHR, driven by
stellar winds and SNe over the last 1600\,yr.

As already mentioned in Sect.~\ref{N4861Havelo}, the northern part of the GEHR
shows a blue- and a red-shifted component in \Ha\ with a velocity offset of
about 30\skms. Figure~\ref{n4861lineprofileshi}, spectrum b only shows some
red-shifted emission at the same position with velocities of about 837\skms,
which corresponds to an offset of about 25\skms\ with respect to the main \HI\
component.

The region west of NGC\,4861 where the three supergiant shells are located
reveals a red-shifted component (see Fig.~\ref{N4861.hi.decomp}, right
panel, Fig.~\ref{n4861lineprofileshi}, spectrum c). The velocities vary
between 860 and 870\skms\ in comparison to the main component of 830 to
840\skms, which is comparable to the ionised gas as the \Ha\ velocity map
shows a red-shifted component of 860 to 870\skms.

At the position of SGS5, a blue-shifted component with an expansion velocity
of 20\skms\ was detected in addition to the main component, which is much
lower than measured in \Ha.
\subsection{\HI\ kinematics of NGC\,4861\,B}
\label{HIcloud}
\begin{figure*}
\centering
\includegraphics[width=\textwidth,viewport= 53 529 580 673,clip=]{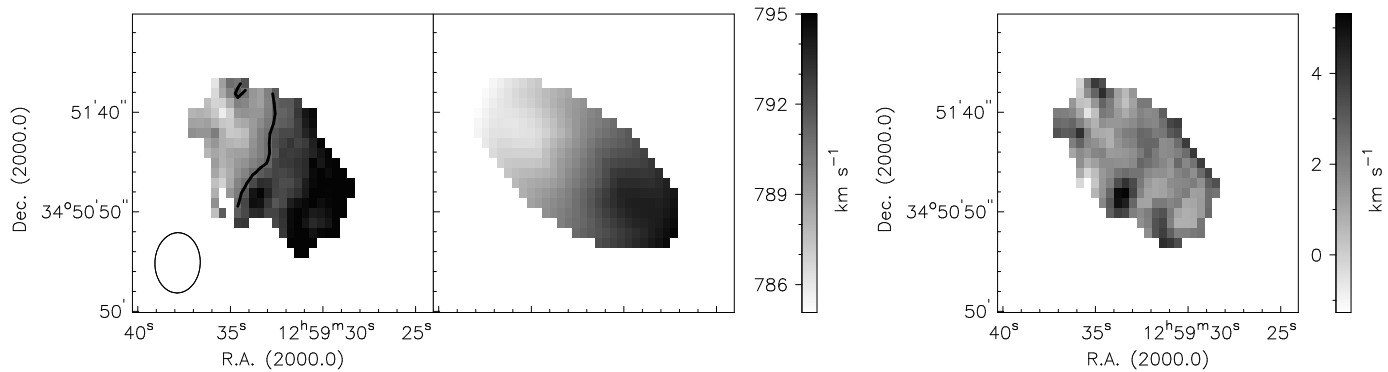}
\caption[A kinematic study of the \HI\ cloud NGC\,4861\,B.]{A kinematic study
  of the \HI\ cloud NGC\,4861\,B. Velocity field (left panel) with a black
  contour at $\vsys=791\skms$, model velocity field (middle panel) and
  residual map (right panel).}
\label{N4861Brot}
\end{figure*}

In order to study the \HI\ kinematics of NGC\,4861\,B, we performed the same
analysis for the \HI\ cloud as for the whole galaxy (see
Sect.~\ref{N4861Hirot}) treating the cloud as an individual system. As we were
limited by both spectral and spatial resolution and as the column density is
quite low, we worked with the values derived by \emph{ellfit} as the
best-fitting parameters. Therefore, the results are just an
estimate. Figure~\ref{N4861Brot} shows the velocity field (left panel), the
model (middle panel) and the residual map (right panel). The velocity field
reveals a regular rotation pattern with a rotation velocity of 4.4\skms. The
inclination is the same as the one of NGC\,4861, but the cloud rotates under a
position angle of 244\degr\ (see Table~\ref{n4861hiparams}). The systemic
velocity was measured to be 791\skms, which is about 60\skms\ lower than
expected from the rotation of the main body.
\section{Discussion}
Our analysis has shown that NGC\,4861 harbours several outflows, both in \Ha\
and \HI. In this section, we want to discuss our detections and to make some
predictions about the fate of the gas.
\subsection{The outflows}
\label{Sectoutflows}
Three prominent outflows were detected which show the same kinematic behaviour
in \Ha\ and \HI. In all cases, the neutral outflowing gas is much more extended
than the ionised gas. This is probably due to the much lower column densities
of the ionised gas, which means that the sensitivity of the FP data is not
high enough to detect it everywhere. We showed in Sect.~\ref{N4861morpho} that
the \HI\ column density (see Fig.~\ref{N4861HI}, upper left panel) is
differently distributed at the positions of the outflows in comparison to the
rest of the galaxy. In the areas of expanding gas the distance between two
neighboured intensity contours becomes much larger. This means that the gas
density is higher in these areas, which fits to the image of gas which is
blown out and is at the same time moving more gas out of the GEHR and
therefore enhancing the density.

From the geometry and the similar expansion velocities the blue- and
red-shifted outflows to the south and to the north of the GEHR probably belong
to the expanding supergiant shell SGS4 detected by \citet{vanEymeren2007}. They
found SGS4 in five highly-resolved spectra with increasing line of sight
velocities of about 110\skms\ blue-shifted and 60\skms\ red-shifted towards the
centre of the GEHR. In \Ha, we could only measure line of sight velocities at
the edges of the GEHR (about 30\skms) because areas of very high intensity
(i.e., the GEHR) are contaminated by a high artificial velocity gradient due
to blurring inside the detector or a saturation problem of the photon counting
system \citep[see][]{vanEymeren2008PhD}. The \HI\ velocity field also shows
expanding gas in the central parts of the GEHR, but only with velocities of
about 25\skms\ blue- and redshifted. This could be a resolution effect as the
GEHR is only represented by a few beams in \HI. Furthermore, the \HI\ has at
least partly been ionised, which is indicated by the lower column density (see
Sect.~\ref{N4861morpho}). Taking all this together, we come to the conclusion
that the fast expanding supergiant shell exists, at least in \Ha.

Outflowing gas was also detected west of NGC\,4861 at the positions of the
supergiant shells. Looking at the \Ha\ image it seems that all three shells
(SGS1, SGS2, SGS3) are connected and form indeed one single huge
shell. However, the kinematic analysis showed that only SGS1 and SGS2 are
expanding, whereas SGS3 shows no significant velocity offset along the line of
sight (see Sect.~\ref{N4861Havelo}). Several scenarios are plausible: all
three shells could have the same origin, but the \HI\ column densities close
to SGS3 were higher and forced the gas to slow down more rapidly. However,
looking at Fig.~\ref{N4861ha+hi}, one can see that SGS3 is located in an area
of lower \HI\ column density than the other two supergiant shells. It is also
possible that SGS3 originates from a different star formation event, which
could be examined by a chemical analysis of the gas. A third option is that
the shell expands perpendicularly to the line of sight.

In the area of SGS5, a faint blue-shifted component with a high velocity
offset of about 80\skms\ was detected in \Ha\ (see
Sect.~\ref{N4861Havelo}). This area was not covered by the echelle spectra
analysed in \citet{vanEymeren2007}, and as the \Ha\ emission is very weak, we
do not want to confirm its existence at this stage. In case it is real
emission, NGC\,4861 becomes an interesting object because it then harbours two
supergiant shells with very high expansion velocities, which requires a very
high energy input and which enhances the chance of a galactic wind. The \HI\
spectra in the area of SGS5 reveal a blue-shifted component with an expansion
velocity of 20\skms, which is far below the offsets measured in \Ha. But
again, this could be a resolution effect.

Where does the gas get its energy from? The outflows close to
the GEHR are most probably driven by the star formation activity within this
\HII\ region. \citet{Barth1994} estimated the age of the star cluster in the
centre of the GEHR to be 4.5\,Myr, which was confirmed later by
\citet{Fernandes2004}. Assuming a constant expansion velocity of 30\skms\ and
a deprojected distance of 545\,pc, we get an estimate for the expansion age of
the blue- and red-shifted outflow south of the GEHR of 1.8\,$\times
10^7$\,yr. For the red-shifted outflow north of the GEHR the expansion age is
even higher with 3.6\,$\times 10^7$\,yr (assumed deprojected distance of
1090\,pc, expansion velocity of 30\skms). This means that either the gas was
ionised by a former star formation event or the expansion velocity has
decreased over time. This result is in good agreement with similar estimates
done for other galaxies \citep{vanEymeren2008}. However, studies by
\citet{Sramek1986} found a deficiency of nonthermal emission in NGC\,4861,
which indicates a lack of supernova remnants and therefore a lack of past star
formation events. This would then mean that the gas was indeed ionised by the
current star formation event and is already loosing kinetic energy. On the
other hand, we probably only miss the high velocity gas due to the technical
problems described above. If we take the expansion velocities of, e.g.,
110\skms\ measured by \citet{vanEymeren2007}, we get an expansion age of
4.8\,$\times 10^6$\,yr at a travel distance of 545\,pc, which is very close to
the age of the star cluster.

We have also looked at the distribution of the hot ionised gas. XMM-Newton data
show that NGC\,4861 has a very luminous X-ray source in the centre of the GEHR
\citep{Stobbart2006}, which would explain the high expansion velocities
detected in the echelle spectra. In the area of SGS5, however, no X-ray
emission can be seen.
\subsection{Outflow or galactic wind?}
\begin{figure}
\centering
\includegraphics[width=.5\textwidth,viewport= 48 516 330 723,clip=]{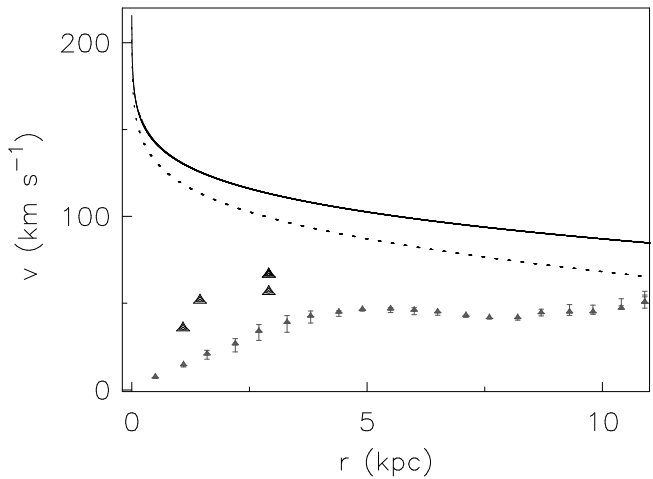}
\caption[A comparison of the escape and expansion velocities.]{Escape velocity
  for a pseudo-isothermal halo of $r_{\rm max}\rm =11\,kpc$ (dotted line) and
  $r_{\rm max}\rm =22\,kpc$ (solid line). The observed rotation curve is
  indicated by small grey triangles. The error bars represent receding and
  approaching side. The expanding gas structures are marked by large black
  triangles.}
\label{N4861esc}
\end{figure}

We now want to compare the expansion velocities of the detected outflows with
the escape velocities of NGC\,4861 in order to make some statements about the
fate of the gas. As shown in \citet{vanEymeren2008PhD}, the pseudo-isothermal
(ISO) halo represents the observed density profiles of dwarf galaxies much
better than the NFW halo, at least in the inner kpcs. As the outflows are
generally close to the dynamic centre, we decided to estimate the escape
velocities from the ISO halo. We followed the same procedure as described in
\citet{vanEymeren2008}. The circular velocity was measured from the rotation
curve to be 46\skms\ (see Table~\ref{n4861hiparams}). In Fig.~\ref{N4861esc}
we plotted the escape velocity for two different halo radii $r_{\rm max}\rm
=11\,kpc$ (dotted line) and $r_{\rm max}\rm =22\,kpc$ (solid line). The lower
value of $r_{\rm max}$ was chosen to equal the size of the \HI\ distribution
so that the corresponding curve is a lower limit for the escape
velocities. The observed rotation curve including receding and approaching
side is indicated by small grey triangles. The expanding gas structures are
marked by large black triangles. We corrected the values for an inclination of
65\degr\ as calculated in Sect.~\ref{N4861HIvelo}, which leads to an increase
in velocity of about 10\%.

As Fig.~\ref{N4861esc} shows, the expansion velocities of all outflows stay
far below the escape velocities, which makes it impossible for the gas to
leave the gravitational potential of the galaxy. However, as discussed in
Sect.~\ref{Sectoutflows}, we probably missed the high velocity parts of the
supergiant shell SGS4 expanding from the GEHR because of the artificial
emission. We did also not include the fast expanding supergiant shell
SGS5. SGS4 is close to the dynamic centre where the escape velocities are
high. SGS5, whose existence first needs to be confirmed by, e.g., obtaining a
slit spectrum, is about 4\,kpc away from the dynamic centre and expands with
80\skms, which means that here, the chance of a galactic wind is strongly
increased.
\subsection{A comparison with NGC\,2366}
\label{Comparison}
We want to compare our results with the results from a previous study of the
irregular dwarf galaxy NGC\,2366 \citep{vanEymeren2008}. NGC\,2366 is very
well suited to be compared to NGC\,4861 because not only their optical
appearances are very similar, but also the parameters describing the kinematic
properties of the neutral gas are similar for both galaxies.

Optical images of NGC\,2366 and NGC\,4861 show that their \Ha\ luminosities
are dominated by a GEHR in the south. At the edge of the star distribution at
the northernmost tip, both galaxies show shell-like structures. In case of
NGC\,2366 we detected an associated star cluster, but in NGC\,4861 it is not
that obvious.

The morphological properties of the neutral gas are different. NGC\,2366 shows
a distribution that is more patchy and a velocity field that is more
perturbed. The main drivers for the distortions are probably the GEHR and the
weak spiral arms. NGC\,4861 shows no evidence for spiral
arms. Nevertheless, this galaxy seems to be more active than NGC\,2366 as it
harbours at least one fast expanding bubble (see Sect.~\ref{Sectoutflows}).

Most of the detected outflows are in the vicinity of the GEHRs, which is to be
expected because the GEHRs are the centres of star formation
activity. NGC\,4861 also shows outflowing gas clearly outside the disc, but
with quite moderate expansion velocities. All outflows were detected in both
\Ha\ and \HI. The 50\skms\ outflow in NGC\,2366, however, was only detected in
\Ha, which is probably an effect of age and energy input. In most cases, a
former star formation event or decreasing expansion velocities are needed to
explain the current position and expansion velocity of the gas
structures.

The main result of our study is that we did not detect any galactic wind in
any of the two galaxies. As already discussed in \citet{vanEymeren2008}, this
is in good agreement with simulations by \citet{MacLow1999} and
\citet{Silich1998}. However, in both galaxies the escape velocity drops
significantly at greater distances from the dynamic centre. In NGC\,4861 we
even detected gas far outside the galactic disc, but it expands with quite
moderate velocities. The gas with high velocity offsets is located in the disc.
\subsection{NGC\,4861\,B}
We obtained a deep \emph{V} band image which confirms the result from
  \citet{Wilcots1996} who did not detect an optical counterpart on a DSS
  image. Furthermore, we could show that the \HI\ cloud seems to be
  kinematically decoupled from the main body as its systemic velocity is
  about 60\skms\ lower and as it is rotating in the opposite direction of
  NGC\,4861. Nevertheless, the distortions in the eastern part of NGC\,4861
  suggest that both systems interact with each other.

We got an idea of the kinematic properties of NGC\,4861\,B by fitting
  ellipses to the \HI\ distribution (see Sect.~\ref{HIcloud}). This means that
  the parameters given in Table~\ref{n4861hiparams} are just
  estimates. Whereas the systemic velocity and the position angle are
  well-defined, the uncertainties of the inclination and therefore the
  inclination corrected rotation velocity and the dynamic mass are
  high. Nevertheless, the velocity field (see Fig.~\ref{N4861Brot}) shows a
  regular rotation pattern and the residual map is smooth. A better
  measurement of the inclination and a higher spatial resolution and
  sensitivity of the \HI\ data are needed in order to investigate the origin
  of this \HI\ cloud.
\section{Summary}
Fabry-Perot interferometric data and \HI\ synthesis observations were used to
get new insights into the morphology and the kinematics of the neutral and
ionised gas components in the nearby irregular dwarf galaxy
NGC\,4861. Additionally, we examined the \HI\ cloud NGC\,4861\,B in more
detail. The most important results are now briefly summarised.

Both gas components show a very similar behaviour. We detected three prominent
outflows, a blue- and a red-shifted outflow in the south of the GEHR with expansion
velocities of about 25\skms\ (\HI) and 30\skms\ (\Ha) together with a blue-
and red-shifted outflow in the north of the GEHR with an expansion velocity of
about 25\skms\ (\HI, only red-shifted component detected) and 30\skms\
(\Ha). These outflows are most probably part of the fast expanding supergiant
shell SGS4 detected in \citet{vanEymeren2007}. Furthermore, a red-shifted
outflow was detected close to the supergiant shells in the west of the tail
with an expansion velocity of about 30\skms\ (\HI\ and \Ha). All outflows are
more extended in \HI\ than in \Ha. A comparison of the expansion velocities
with the escape velocity of the galaxy using the pseudo-isothermal halo model
revealed that the gas stays gravitationally bound. This is in good agreement
with a previous study of the dwarf galaxy NGC\,2366 \citep{vanEymeren2008}.

A deep \emph{V}-band image suggests that the \HI\ cloud NGC\,4861\,B has no
optical counterpart as we did not detect any star association down to a
3$\sigma$ detection limit of 27.84\,mag\,arcsec$^{-2}$ in
\emph{V}. Nevertheless, it shows a regular rotation. From our results, it is
not possible to define its origin.
\begin{acknowledgements}
The authors would like to thank Eric Wilcots for providing the VLA \HI\ data
and the \emph{R}-band image of NGC\,4861, for stimulating discussions and the
invitation to Madison. We also thank the referee, Kambiz Fathi, whose 
persistent feedback helped us to deepen our knowledge about handling and 
interpreting line profiles.\\
This work was partly supported by the Deutsche Forschungsgesellschaft (DFG)
under the SFB 591, by the Research School of the Ruhr-Universit\"at Bochum, by
the Australia Telescope National Facility, CSIRO, and by the DAAD. It is
partly based on observations collected at the Observatoire de
Haute-Provence and at the Centro Astron$\rm\acute{o}$mico Hispano
Alem$\rm\acute{a}$n (CAHA) at Calar Alto, operated jointly by the Max-Planck
Institut f\"ur Astronomie and the Instituto de Astrof\'{i}sica de
Andaluc\'{i}a (CSIC). It is also partly based on archival VLA data of the
National Radio Astronomy Observatory. The NRAO is a facility of the National
Science Foundation operated under cooperative agreement by Associated
Universities, Inc. We made extensive use of NASA's Astrophysics Data System
(ADS) Bibliographic Services and the NASA/IPAC Extragalactic Database (NED)
which is operated by the Jet Propulsion Laboratory, California Institute of
Technology, under contract with the National Aeronautics and Space
Administration.
\end{acknowledgements}
\bibliographystyle{aa}
\bibliography{11766}
\appendix
 \section{\Ha\ image and extension of the catalogue of ionised gas structures}
\label{App5}
\begin{figure*}
\centering
\includegraphics[width=.8\textwidth,viewport= 50 524 271 731,clip=]{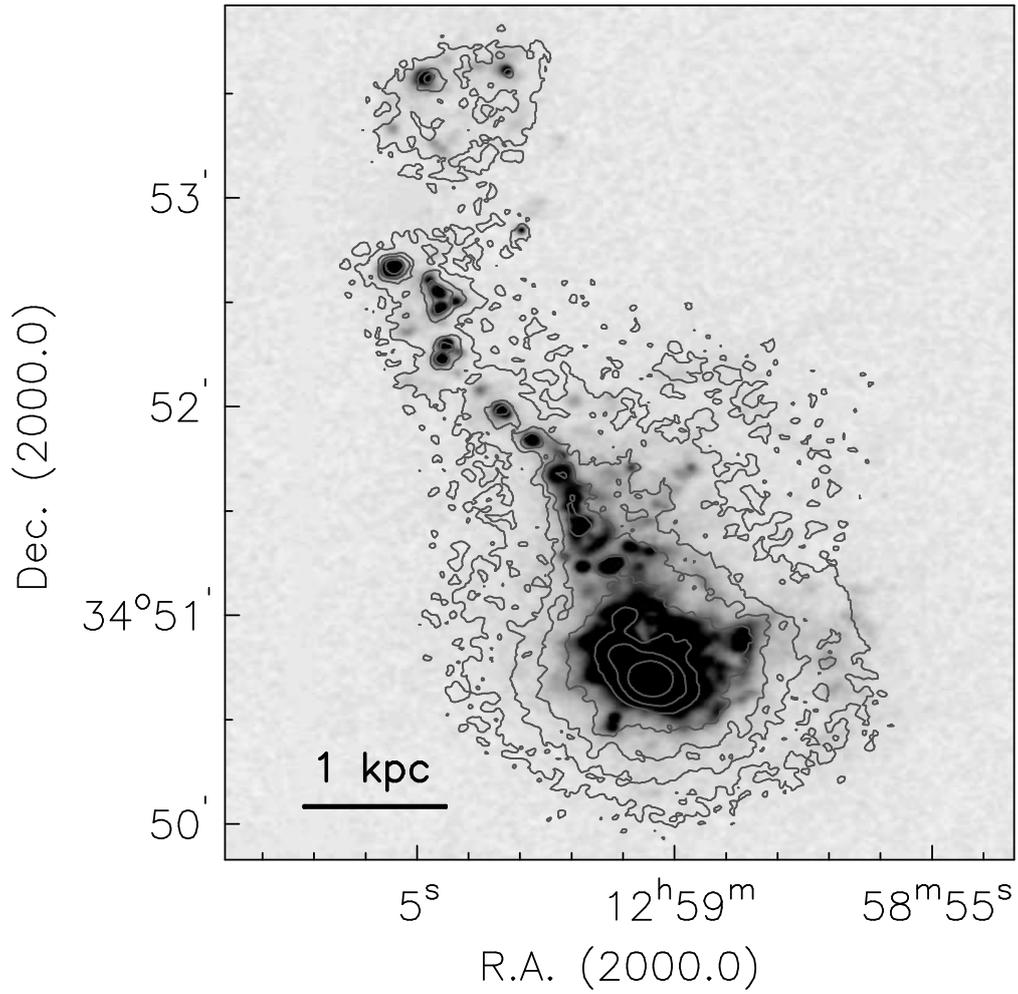}
\caption[Continuum-subtracted \Ha\ image of NGC\,4861.]{Continuum-subtracted
  \Ha\ image of NGC\,4861. The contrast is chosen in a way to demonstrate the
  small-scale structures. Overlaid in grey are the FP \Ha\ intensity
  contours at 0.5 (3$\sigma$), 5, 10, 20, 50, 100, and 300 (arbitrary units).}
\label{N4861ha}
\end{figure*}

Here, the continuum-subtracted \Ha\ image is again presented in an enlarged
version and with a different contrast as in Fig.~\ref{N4861r+ha} to emphasise
the small-scale structures (see Fig.~\ref{N4861ha}). For a comparison,
  the FP \Ha\ intensity contours are overlaid in grey.
\end{document}